\documentclass[aps,amsmath,amssymb,twocolumn,groupedaddress,showpacs]{revtex4}

\usepackage{graphicx}
\usepackage{dcolumn}
\usepackage{bm}
\usepackage{color}

\begin{document}

\title{S-wave and p-wave scattering in a cold gas of Na and Rb atoms}

\author{H. Ouerdane}
\affiliation{CIMAP, CEA-CNRS-ENSICAEN 6252, BP 5133, F-14070 Caen, France.}
\author{M. J. Jamieson}
\affiliation{Department of Computing Science, University of Glasgow, Glasgow G12 8QQ, United Kingdom}
\email{mjj@dcs.gla.ac.uk}

\begin{abstract}
Using improved experimentally based $X{}^1\Sigma^+$ and $a{}^3\Sigma^+$ molecular potentials of NaRb, we apply the variable phase method to compute new data for low energy scattering of $^{23}$Na atoms by  $^{85}$Rb atoms and $^{87}$Rb atoms. These are the scattering lengths and volumes, numbers of bound states and effective ranges, which we use to obtain the low energy spin-change cross section as functions of the system temperature and the isotope masses. From an analysis of the contributions of s-wave and p-wave scatterings to the elastic cross section we estimate temperatures below which only s-wave scattering is dominant. We compare our quantal results to data obtained from the semiclassical approximation. We supply evidence for the existence of a near zero energy  p-wave bound state supported by the singlet molecular potential.
\end{abstract}


\maketitle

\section{Introduction}

The properties of ultracold trapped ensembles of atoms are governed by low-energy collisions that occur within such systems \cite{WEI99}. Of particular interest is the study of heteronuclear alkali-metal dimers because it is now possible to generate dual species Bose-Einstein condensates (BEC) and produce Bose-Fermi ensembles from such binary mixtures \cite{TRU01,SCH01,ROA02,MOD02}. At sufficiently low temperatures a few scattering parameters suffice to characterise collisions between two atoms of different species. These are the s-wave scattering length, $a_{\rm s}$, the p-wave scattering volume, $a_{\rm p}$, and the effective range, $R_{\rm eff}$, for scattering via the singlet and triplet ground states molecular potentials; but their accurate computation remains non-trivial.

It is important to obtain precise scattering data since they give valuable information on the processes that allow or not the production of a stable condensate. In studies of the condensate it is important to know the low temperature elastic scattering cross sections, which control the thermalisation of the atoms. The spin-change cross sections are also important. Atoms interacting through the triplet state of NaRb remain in a magnetic trap but those in the singlet state leave. Thus spin-change is an important trap loss mechanism. We present calculations of the scattering lengths and volumes and the effective ranges that suffice to quantify low energy elastic scattering, and of the spin-change cross section, which quantifies trap loss. We note that in a full treatment of stability of the condensate the hyperfine structure must be accounted for \cite{ESR97,BUR97} but in such a treatment the scattering data that we present are needed.

The two main processes that dictate the properties of a trapped cold atomic gas in thermal equilibrium, are the elastic collision and the spin change collision. The elastic collision process is essentially of s-wave nature if the temperature is sufficiently low. The spin change collision is an inelastic process during which the spins up and down ($s$ and $s'$) of two scattering atoms, $A_1$ and $A_2$, are exchanged:

$$
A_1(s) + A_2(s') \longrightarrow A_1(s') + A_2(s)
$$

\noindent In addition to the stability criterion that has to be satisfied to obtain a BEC, it is also important for practical purposes to ensure that the system allows efficient evaporative cooling. This is possible if the ultracold s-wave elastic collisions dominate the inelastic spin change collisions, i.e. if the related cross sections satisfy $\sigma_{\rm el} \gg \sigma_{\rm sc}$. For a given pair of atoms of different species in the singlet or triplet state ($\rm s$/$\rm t$), the low energy elastic scattering cross section is proportional to the square of the s-wave scattering length, $a_{\rm s}^{({\rm s}/{\rm t})}$ and the spin change cross section, to the square of the difference of the two scattering lengths, $a_{\rm s}^{({\rm s})} - a_{\rm s}^{({\rm t})}$. It is of interest to estimate an upper bound to the temperature range in which only s-wave scattering is significant and to investigate the dependence of the spin-change cross section on the atomic masses and the temperature.

Previously we demonstrated the utility of the variable phase method to circumvent numerical technical difficulties and to accelerate convergence of the accumulated scattering parameters \cite{OUE03,OUE04}. However the reliability of computed data also depends on the accuracy of the molecular potentials; it is possible that, within its error bounds, a molecular potential supports a near zero-energy bound state which makes the computed scattering length extremely sensitive to any change in the potential. Cold collisions of Na and Rb atoms are interesting in photoassociation and trap loss experiments and we evaluated scattering lengths and volumes, and numbers of bound states computed using various potentials ~\cite{OUE04}. We investigated the mass dependence of our results, which were compared to the data published by Weiss \emph{et al.} \cite{WEI03}. It has been suggested by C\^{o}t\'{e} {\it et al.} \cite{COT05} that our calculations were tentative because the potentials need to be specified more precisely. The new potentials for NaRb of Pashov {\it et al.} \cite{PAS05} represent an improvement over those that were available at the time of our calculations \cite{OUE04} but, contrary to their suggestion \cite{PAS05}, the confusion of units in the work of Weiss \emph{et al.} \cite{WEI03} does not compromise the reliability of our earlier calculations since we did not use but merely quoted their results, which were corrected later \cite{WEI04}.

For the present work, we constructed molecular potentials using the improved short range data of Pashov {\it et al.} \cite{PAS05}. The exchange potential is given by the expression of Smirnov and Chibisov \cite{SMI65} in the formulation used by Weiss \emph{et al.} \cite{WEI03}. Smirnov and Chibisov \cite{SMI65} showed that the exchange potential between two heteronuclear atoms in their ground $n{\rm s}^2{\rm S}$ states can be approximated by the series (which having a single term is exact for homonuclear atoms) 

\begin{equation}\label{exch}
V_{\rm E}(R)= -\sum_{n=0}^N \left[J_n R^n \frac{(\alpha'-\alpha)^n}{n!}\right] R^{\gamma} \exp (-\beta R),
\end{equation}

\noindent where $V_{\rm E}(R)$ hartree is the exchange potential at $R$ bohr, $\alpha^2/2$ hartree is the ionisation energy of an atom, $\beta=\alpha+\alpha'$ and $\gamma=2/\alpha+2/\alpha'+1/(\alpha+\alpha')-1$; $J_n$ is calculated from the asymptotic behaviours of the wavefunctions of the valence electrons of the atoms and it also depends on quadratures over functions with branch points. We evaluated the values of $J_n$, shown in Table \ref{tab:tableJ}, for alkali metal atom pairs using the calculations of Marinescu and Dalgarno \cite{MAR96} for the asymptotic wavefunctions of the valence electrons. The definition of $J_n$ is half that used by Smirnov and Chibisov \cite{SMI65}. We used experimental ionisation energies \cite{NISTJ}. The values of $J_0$, $J_1$ and $J_2$ for NaRb agree with those of Weiss {\it et al.} \cite{WEI03}. We note that Zemke and Stwalley \cite{ZEM99} proposed a representation of the exchange interaction equivalent to a single term of equation (1), $-CR^\alpha \exp(-\beta R)$ with $\alpha$ and $\beta$ to be determined. For consistency in comparison with the calculation of Weiss {\it et al.} we used their formulation {\it i.e.} equation (1).

Long range molecular potentials of alkali-metal atoms are usually well characterised by three van der Waals dispersion coefficients $C_6$, $C_8$ and $C_{10}$ \cite{STA85}; we took the values computed by Derevianko \emph{et al.} \cite{DER01,DER03} and we  added higher order inverse powers where we used the coefficients beyond $C_{10}$ computed by Mitroy and Bromley \cite{MIT05}. The short range potentials of Pashov {\it et al.} \cite{PAS05}, being experimentally based, account for adiabatic, diabatic and relativistic effects. The long range dispersion potentials do not account for such effects but we can consider adiabatic corrections. Dalgarno and McCarroll \cite{DAL56} showed that allowance for the adiabatic coupling between the nuclear and the electronic motion changes the van der Waals coefficients in proportion by amounts of order of the inverse of the reduced mass, but in these calculations these corrections would be smaller than 0.01\% and therefore we do not include them.

In Section 2, we give an overview of the theory and discuss the semiclassical approximation to the p-wave scattering volume. We show how we produce estimates of the temperature range in which only s-wave scattering is significant from a simplified analysis of the s-wave and p-wave cross sections; and we derive an expression for the low energy spin-change cross section. In Section 3, we present and discuss numerical data.

\section{Theory}

\subsection{Low energy scattering}

The accumulated s-wave phase, $\delta_k(R)$, for wavenumber $k$ of relative motion, satisfies the
differential equation in the nuclear separation $R$ of the Riccati type \cite{CAL67}:

\begin{equation} \label{eq1}
  \frac{\displaystyle {\rm d}\delta_k(R)}{\displaystyle {\rm d}R} = -k^{-1} V(R) \sin^2\left[kR+\delta_k(R)\right],
\end{equation}

\noindent where $V(R)$ is $2\mu/\hbar^2$ times the interaction potential and $\mu$ is the reduced mass of the dimer. The phase shift is the limit of the accumulated phase at infinite separation and we write
it as $\delta_k=\delta_k(\infty)$; it suffers no mod[$\pi$] ambiguity. The number of s-wave bound states supported by the potential is  given by Levinson's theorem \cite{CAL67,LEV49} as $\pi N_{\rm b} = \lim_{k\rightarrow 0}\delta_k$.

In effective range theory \cite{LEV63,HIN71,MOT65} the low energy expansion of the tangent of the s-wave scattering phase shift yields the equations satisfied by the accumulated scattering length, $a_{\rm s}(R)$, and a volume, $b_{\rm s}(R)$, that is the coefficient that appears to third order in the expansion of $\tan \delta_k(R)$ for small values of $k$:

\begin{eqnarray} \label{eq2a}
\frac{\displaystyle {\rm d}a_{\rm s}(R)}{\displaystyle {\rm d}R} & = & \left[R - a_{\rm s}(R)\right]^2 V(R)\\
\label{eq2c}
\frac{\displaystyle {\rm d}b_{\rm s}(R)}{\displaystyle {\rm d}R} & = &\!\! \left\{\!\frac{\displaystyle \left[R-a_{\rm s}(R)\right]^4}{\displaystyle 3}\!-\!2\left[R-a_{\rm s}(R)\right]b_{\rm s}(R)\!\right\}\!\!V(R)
\end{eqnarray}

\noindent The accumulated effective range, $R_{\rm eff}(R)$, is related to the accumulated scattering length and the volume by:

\begin{equation}\label{eq2d}
R_{\rm eff}(R) = \frac{\displaystyle 2a_{\rm s}(R)}{\displaystyle 3} - \frac{\displaystyle 2b_{\rm s}(R)}{\displaystyle \left[a_{\rm s}(R)\right]^2}.
\end{equation}

\noindent Eq. (\ref{eq2c}), satisfied by $b_{\rm s}(R)$, is coupled to Eq.~(\ref{eq2a}) and has a closed form solution \cite{CAL67,OUE04b} but the solution of Eq.~(\ref{eq2c}) is difficult to compute because of the divergence problems that reflect the presence of poles in the accumulated scattering length $a_{\rm s}(R)$. We have discussed this in detail in Ref.~\cite{OUE04b}, where an analytical first order long range correction to the effective range is given. In Refs.~\cite{OUE03,OUE04} we showed how upper and lower bounds to the scattering length and scattering volume can be obtained and gave expressions of improved approximations to the scattering parameters as linear combinations of those bounds.

In the following we mean by the radial wave function, $R$ times the actual radial wave function and, by the Schr\"odinger equation, the equation that is satisfied by this radial wave function (that contains no first derivative). In our previous analysis \cite{OUE04} we related $l$-wave scattering to s-wave scattering using a relation given by Calogero \cite{CAL67}; here we present a simplified version for p-wave scattering. Making the substitution $Z=R^3/3$ and writing the zero energy radial wave function as $R^{-1}\phi(Z)$, we obtain, from the p-wave Schr\"odinger equation, an s-wave Schr\"odinger equation that is satisfied by $\phi(Z)$ but with the potential divided by $({\rm d}Z/{\rm d}R)^2$. The asymptotic solution, suitably normalized, is $Z-a$ where $a$ is a volume analogous to the scattering length. Application of variable phase theory with $\phi(Z)=Z-a(Z)$, where $a(Z)$ is analogous to the accumulated scattering length, yields an  equation like Eq. (\ref{eq2a}). However the function $R^{-1}(Z-a)$ is the same as the function $R^2/3-a/R$, which is the suitably normalised asymptotic p-wave function \cite{GUT84}. Hence the s-wave scattering quantity $a$ for the potential divided by $({\rm d}Z/{\rm d}R)^2$ {\emph is} the p-wave scattering volume. On writing the equation in $Z$ that is equivalent to  equation (\ref{eq2a}) we find the following equation that is satisfied by the accumulated p-wave scattering volume $a_{\rm p}(R)$: 

\begin{equation}
\label{eq2b}
\frac{\displaystyle {\rm d}a_{\rm p}(R)}{\displaystyle {\rm d}R} = \left[\frac{\displaystyle R^3}{\displaystyle 3} - a_{\rm p}(R)\right]^2 R^{-2}V(R).
\end{equation}

Recently Dickinson \cite{DIC08} derived a semiclassical formula for the scattering volume. We can derive his formula directly from the relation

\begin{equation}
\int_{R_0}^\infty \sqrt{-V(R)}~{\rm d}R =
\int_{Z_0}^\infty \sqrt{-V(Z)}~\frac{{\rm d}R}{{\rm d}Z}~{\rm dZ}
\end{equation}

\noindent where $V(R_0)=0$ and $Z_0={R_0}^3/3$, and the semiclassical s-wave formula of Gribakin and Flambaum \cite{GRI93}. With $Z=R^3/3$, the leading term of the long range potential is $-(C_6/27\sqrt[3]{3})/Z^{10/3}$ where C$_6$ is the van der Waals dispersion coefficient (in terms of $R$ the leading long range term is $-C_6/R^6$). Substituting $\alpha={\rm C}_6/27\sqrt[3]{3}$ and $n=10/3$ into the formula of Gribakin and Flambaum we obtain the semiclassical formula of Dickinson. The Schr\"odinger equation for $l$-wave scattering can be reduced to an s-wave scattering equation by the transformation $Z=R^{2l+1}/(2l+1)$ \cite{OUE04,CAL67}. This transformation and the formula of Gribakin and Flambaum can be used in the manner described above to find a generalisation of the semiclassical formula for the quantity analogous to the scattering length, provided it exists, for higher angular momenta and potentials that behave asymptotically as $R^{-n}$. Below we compare our quantal results for the scattering volume with those predicted by Dickinson's formula.

\subsection{Temperature range}

Knowledge, even approximate, of an upper bound for the temperature range, $T$, in which s-wave scattering dominates the elastic collisions in the trapped gas of ultracold atoms, is useful. Here, we show how estimates can be found from a simple analysis of the s-wave and p-wave scattering cross sections. For small values of the wavenumber $k$ the s-wave scattering contributes 

\begin{equation}\label{eq3}
\sigma_{\rm el,s} = 4\pi a_{\rm s}^2 \left(1 - k^2a_{\rm s}^2 + k^2 R_{\rm eff}a_{\rm s}\right)
\end{equation}

\noindent to the elastic cross  section. If the gas is sufficiently cold that $k|a_{\rm s}|\ll 1$ and $k^2 R_{\rm eff}|a_{\rm s}|\ll 1$, where $k$ now denotes the thermally averaged wavenumber, then the cross section is not influenced by the effective range and is also well approximated by $\sigma_{\rm el,s} \approx 4\pi a_{\rm s}^2$. If the second inequality, $k^2 R_{\rm eff}|a_{\rm s}|\ll 1$, is satisfied then so is the first (unless the effective range is pathologically large). In this case, the temperature $T$ satisfies $T \ll \hbar^2/2\mu k_{\rm B}R_{\rm eff}|a_{\rm s}|$. In estimating an upper bound to the temperature, let us require that the value of $k$ be smaller than 1\% of $1/\sqrt{R_{\rm eff}|a_{\rm s}|}$. The contribution of p-wave scattering to the elastic cross section is given by $\sigma_{\rm el,p} = 12\pi k^4a_{\rm p}^2$, which can be rewritten as $\sigma_{\rm el,p} = 4\pi a_{\rm s}^2 \times 3k^4a_{\rm s}^4 \left(a_{\rm p}/a_{\rm s}^3\right)^2$. If $k$ satisfies the above criteria, which it does if $T \ll \hbar^2/2\mu k_{\rm B}R_{\rm eff}|a_{\rm s}|$, then $\sigma_{\rm el,p}/\sigma_{\rm el,s} \ll 1$ and p-wave scattering can be neglected.

\subsection{Spin change cross section}

The spin change cross section is \cite{DAL61}:

\begin{equation}\label{sccs}
\sigma_{\rm sc} = \frac{\displaystyle 4\pi}{\displaystyle k^2} \sum_{l} (2l+1) \sin^2\left[\delta_{k,l}^{({\rm s})} - \delta_{k,l}^{({\rm t})}\right],
\end{equation}

\noindent where the $\delta_{k,l}^{({\rm s})}$ and $\delta_{k,l}^{({\rm t})}$ denote the phase shifts of the partial wave with angular momentum $l$, in the singlet and triplet states respectively (note that for the s-wave case, in Eq.~(\ref{eq1}), neither the $l$-dependence of the phase shift nor the spin configuration, were made explicit).

In the ultra-low energy limit ($k\approx 0$), only the partial waves with angular momenta $l=0$ and $l=1$ contribute significantly to the scattered wave function, the contributions from higher angular momenta being negligible at very low temperature \cite{MOT65,BLA48,BLA49,BET49,HAD02}. The low energy expansion of the s-wave phase shifts, in the singlet/triplet states, is:

\begin{equation}\label{s-phase}
\delta_{k,0}^{({\rm s}/{\rm t})} = \pi N_{{\rm b},0}^{({\rm s}/{\rm t})} - ka_{\rm s}^{({\rm s}/{\rm t})} 
+ k^3b_{\rm s}^{({\rm s}/{\rm t})},
\end{equation}

\noindent and

\begin{equation}\label{p-phase}
\delta_{k,1}^{({\rm s}/{\rm t})} = \pi N_{{\rm b},1}^{({\rm s}/{\rm t})} - k^3 a_{\rm p}^{({\rm s}/{\rm t})}
\end{equation}

\noindent for the p-wave phase shifts. $N_{{\rm b},0}^{({\rm s}/{\rm t})}$ and $N_{{\rm b},1}^{({\rm s}/{\rm t})}$ are the numbers of s-wave and p-wave bound states. The volume parameter $b_{\rm s}^{({\rm s}/{\rm t})}$ is related to the s-wave scattering length and effective range through equation (\ref{eq2d}). Note that in the limit $k\rightarrow 0$, Levinson's theorem is recovered from Eqs.~(\ref{s-phase}) and (\ref{p-phase}).

Now, inserting Eqs.~(\ref{s-phase}) and (\ref{p-phase}) into Eq.~(\ref{sccs}), we obtain the following expression for the spin-change cross section, in the low energy limit, up to second order in relative motion energy (i.e. up to $k^4$):

\begin{eqnarray}\label{sccs2}
\sigma_{\rm sc} & = & 4\pi\left\{ \left[a_{\rm s}^{({\rm s})} - a_{\rm s}^{({\rm t})}\right]^2 + 2\left[a_{\rm s}^{({\rm s})} - a_{\rm s}^{({\rm t})}\right] \left[b_{\rm s}^{({\rm t})} - b_{\rm s}^{({\rm s})}\right]k^2\right.
\nonumber\\ 
&&{}+ \left.\left(\left[b_{\rm s}^{({\rm t})} - b_{\rm s}^{({\rm s})}\right]^2 + 3 \left[a_{\rm p}^{({\rm s})} - a_{\rm p}^{({\rm t})}\right]^2\right)k^4\right\}.
\end{eqnarray}

The first term of the right hand side of equation~(\ref{sccs2}), $\sigma_{\rm sc}^0=4\pi \left[a_{\rm s}^{({\rm s})} - a_{\rm s}^{({\rm t})}\right]^2$, represents the zeroth order approximation of the spin change cross section in the low energy limit, and it is temperature independent. The spin-change cross section, $\sigma_{\rm sc}$, is studied as a function of isotope mass and cold atom system temperature, in the next section.

\section{Results and discussion}

First, in Table \ref{tab:tableJ}, we give the coeffients $\alpha$ and $J_n$, appearing in Eq.~(\ref{exch}), computed for alkali dimers. The masses that we used in our calculations of the scattering data are 22.98976967 for ${}^{23}\mbox{Na}$, 84.9117893 for
${}^{85}\mbox{Rb}$ and 86.9091835 for ${}^{87}\mbox{Rb}$ (in atomic mass units) \cite{NIST}.

The scattering data given in Table \ref{tab:table1} are calculated from the potentials in Ref.~\cite{PAS05}.
A useful check on the reliability of our numerical procedures is the agreement of our s-wave scattering lengths with those of Pashov {\it et al.} \cite{PAS05} for both isotopomers in the singlet and triplet states; new data that we provide are the p-wave scattering volumes, s-wave effective ranges and numbers of bound states. Table \ref{tab:table1} shows that the values of the scattering volumes are close to those predicted by Dickinson's relation between the scattering length and the scattering volume (Eq.~(21) of Ref.~\cite{DIC08}) except for the $^{87}$Rb isotope where the quantal and semiclassical values, $a_{\rm p}$ and $a_{\rm p}^{\rm SC}$, are not so close themselves. Note that both values remain large in comparison to the other volumes, which, together with a change of sign, possibly indicates the presence of a p-wave resonance for the $^{23}$Na - $^{87}$Rb cold collision in the singlet state. The presence of a resonance, which is further investigated below, explains the sensitivity of the scattering data to the semiclassical approximation. 

Despite the sensitivity of the computed scattering data to minor changes in the molecular potentials, we find that influence of the van der Waals terms with coefficients $C_{11}$ to $C_{16}$ is negligible: they contribute less than 1\% to the computed scattering parameters. This is consistent with the conclusion of Mitroy and Bromley \cite{MIT05} that, because of cancellation, the contribution of the dispersion terms $-C_n/R^n$ might be neglected when $n>10$  if the accuracy of the van der Waals coefficients is of the order of 1\%. The temperatures below which only s-wave scattering is significant are all smaller than 100 nK; thus p-wave scattering can be neglected only in extremely cold Na and Rb gases, particularly for gases containing the ${}^{85}\mbox{Rb}$ isotope.

\begin{table}
\caption{$\alpha$ and $J_n$ (atomic units).}
\label{tab:tableJ}
\begin{tabular}{lccccccc}
\hline\hline
   &          & ~ &   ~    & ~	    & ~       & ~       & ~       \\
\multicolumn{1}{c}{} &
\multicolumn{1}{c}{$\alpha$} &
\multicolumn{6}{c}{$10^4 \times J_n$} \\
   &          & ~ &   ~    & ~	    & ~       & ~       & ~       \\
   &          & $n$ & Li   & Na	    & K       & Rb      & Cs      \\
   &          & ~ &   ~    & ~	    & ~       & ~       & ~       \\
\cline{3-8}
   &          & ~ &   ~    & ~	    & ~       & ~       & ~       \\
Li &  0.62951 & 0 & 172.85 & 152.22 & 89.522  & 80.350  & 63.424  \\
   &   	      & 1 &        & 1.4498 & 3.9634  & 4.2131  & 4.2858  \\
   &   	      & 2 &        & 21.500 & 12.475  & 11.184  & 8.8276  \\
   &   	      & 3 &        &        & 1.3507  & 1.4304  & 1.4460  \\
   &   	      & 4 &        &        & 4.0052  & 3.5883  & 2.8334  \\
   &          & ~ &   ~    & ~	    & ~       & ~       & ~       \\
\cline{3-8}
   &          & ~ &   ~    & ~	    & ~       & ~       & ~       \\
Na &  0.61459 & 0 &        & 134.13 & 79.050  & 70.986  & 56.082  \\
   &   	      & 1 &        &        & 2.7489  & 3.0487  & 3.2593  \\
   &   	      & 2 &        &        & 10.900  & 9.7686  & 7.7056  \\
   &   	      & 3 &        &        & 0.93017 & 1.0275  & 1.0911  \\
   &   	      & 4 &        &        & 3.4689  & 3.1046  & 2.4470  \\
   &          & ~ &   ~    & ~	    & ~       & ~       & ~       \\
\cline{3-8}
   &          & ~ &   ~    & ~	    & ~       & ~       & ~       \\
K  &  0.56483 & 0 &        &        & 46.942  & 42.226  & 33.466  \\
   &   	      & 1 &        &        &         & 0.34607 & 0.78409 \\
   &   	      & 2 &        &        &         & 5.6200  & 4.4215  \\
   &          & ~ &   ~    & ~	    & ~       & ~       & ~       \\
\cline{3-8}
   &          & ~ &   ~    & ~	    & ~       & ~       & ~       \\
Rb &  0.55409 & 0 &        &         &        & 37.999  & 30.138  \\
   &   	      & 1 &        &        &         &         & 0.45928 \\
   &   	      & 2 &        &        &         &         & 3.9529  \\
   &          & ~ &   ~    & ~	    & ~       & ~       & ~       \\
\cline{3-8}
   &          & ~ &   ~    & ~	    & ~       & ~       & ~       \\
Cs &  0.53497 & 0 &        &        &         &         & 23.936  \\
   &          & ~ &   ~    & ~	    & ~       & ~       & ~       \\
\hline\hline
\end{tabular}
\end{table}

We turn now to the study of the spin change cross section, whose main contribution comes from $\sigma_{\rm sc}^0$. We obtain: $\sigma_{\rm sc}^0 = 1.25\times 10^6~\mbox{bohr}^2$ and $\sigma_{\rm sc}^0 = 1.91\times 10^4~\mbox{bohr}^2$ for the ${}^{23}\mbox{Na}{}^{85}\mbox{Rb}$ and ${}^{23}\mbox{Na}{}^{87}\mbox{Rb}$ isotopes respectively. Considering the values of the elastic cross sections given in Table \ref{tab:table1}, the first important observation is that the criterion $\sigma_{\rm el} \gg \sigma_{\rm sc}$ is not fully satisfied for both isotopomers: stability of the dual species ${}^{23}\mbox{Na}{}^{85}\mbox{Rb}$ and ${}^{23}\mbox{Na}{}^{87}\mbox{Rb}$ condensate against the inelastic spin change process can hardly be achieved. It is also interesting to note that the spin change cross section is also very sensitive to a variation of the atomic mass since it is two orders of magnitude larger for the lighter system than it is for the heavier one. This is coherent with the fact that because of the lighter mass of the atom ${}^{85}\mbox{Rb}$, the larger spatial extension of its wavefunction enhances the electron spin exchange during the scattering act.

\begin{table}
\caption{S-wave scattering lengths $a_{\rm s}$, elastic scattering cross sections $\sigma_{\rm el,s}$, p-wave volumes $a_{\rm p}$ (quantal) and $a_{\rm p}^{\rm SC}$ (semiclassical), effective ranges $R_{\rm eff}$, s-wave and p-wave numbers of bound states $N_{\rm b,0}$ and $N_{\rm b,1}$, and temperatures $T$.}
\label{tab:table1}
\begin{tabular}{ccccc}
\hline\hline
\rule{0pt}{4ex}
&\multicolumn{2}{c}{$X{}^1\Sigma^+$}&\multicolumn{2}{c}{$a{}^3\Sigma^+$}\\
\rule{0pt}{4ex}
 ~ & ${}^{23}\mbox{Na}{}^{85}\mbox{Rb}$ & ${}^{23}\mbox{Na}{}^{87}\mbox{Rb}$ & ${}^{23}\mbox{Na}{}^{85}\mbox{Rb}$ & ${}^{23}\mbox{Na}{}^{87}\mbox{Rb}$  \\ \hline
~&~&~&~&~\\
$a_{\rm s}$ (bohr)  	&$\phantom{-}$396 &109 &81 &70\\
~&~&~&~&~\\
$\sigma_{\rm el,s}$ ($\mbox{bohr}^2$) &$\phantom{-}$1.97$\times 10^6$ & 1.49$\times 10^5$ & 8.25$\times 10^4$ & 6.16$\times 10^4$\\
~&~&~&~&~\\
$a_{\rm p}$ ($\mbox{bohr}^3$)  &$- 4.30\times 10^5$ &$7.0\times 10^6$ &$3.10\times 10^5$ &$1.30\times 10^5$\\
~&~&~&~&~\\
$a_{\rm p}^{\rm SC}$ ($\mbox{bohr}^3$)  &$- 4.25\times 10^5$ &$1.38\times 10^7$ &$3.15\times 10^5$ &$1.31\times 10^5$\\
~&~&~&~&~\\
$R_{\rm eff}$ (bohr) &$\phantom{-}$31 &58 &84 &113\\
~&~&~&~&~\\
$N_{\rm b,0}$ 	&$\phantom{-}$83 &83 &23	&23 \\
~&~&~&~&~\\
$N_{\rm b,1}$ 	&$\phantom{-}$82 &83 &23	&23 \\
~&~&~&~&~\\
$T$ (nK) &$\phantom{-}$53 &75 &70 &60\\
~&~&~&~&~\\
\hline\hline
\end{tabular}
\end{table}

\begin{figure}
\resizebox{1.0\columnwidth}{!}{\includegraphics*{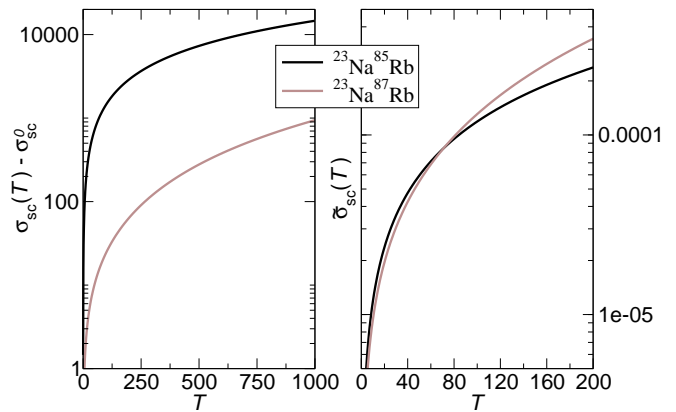}}
\caption{Variations of $\sigma_{\rm sc}(T)-\sigma_{\rm sc}^0$ ($\mbox{bohr}^2$) and ${\tilde \sigma_{\rm sc}}(T)$ with temperature $T$ (K). See text.}
\label{fig:eps2}
\end{figure}

On the left panel of Fig. \ref{fig:eps2}, the variation of $\sigma_{\rm sc}(T)-\sigma_{\rm sc}^0$ is shown for temperatures up to the $\mu$K domain. It is obvious that in this temperature range the first and second order corrections contribute only a little to the total spin change cross section. These corrections remain larger in the case of the lighter isotopomer for all temperatures in the range studied. However, if we define a standardized spin change cross section as ${\tilde \sigma_{\rm sc}}(T) = \left[\sigma_{\rm sc}(T)-\sigma_{\rm sc}^0\right]/ \sigma_{\rm sc}^0$, shown on the right panel of Fig. \ref{fig:eps2}, we find that for temperatures close to 60 K to 70 K, the behavior of the corrections changes: the amplitude of $\tilde \sigma_{\rm sc}(T)$ for ${}^{23}\mbox{Na}{}^{87}\mbox{Rb}$ becomes larger than it is for ${}^{23}\mbox{Na}{}^{85}\mbox{Rb}$. This is due to the fact that the last term of the right hand side of Eq.~(\ref{sccs2}), proportional to the square of the difference of the p-wave volumes, becomes relatively important (compared to the other first- and second-order correction terms) as the temperature increases; specially considering that the magnitude of the p-wave volume of ${}^{23}\mbox{Na}{}^{87}\mbox{Rb}$ in the singlet state, is particularly large compared to the other volumes. Note that this finding is coherent with the simple analysis of the s-wave and p-wave elastic cross sections we presented above: p-wave scattering becomes non-negligible in this temperature range.

\begin{figure}
\resizebox{1.0\columnwidth}{!}{\includegraphics*{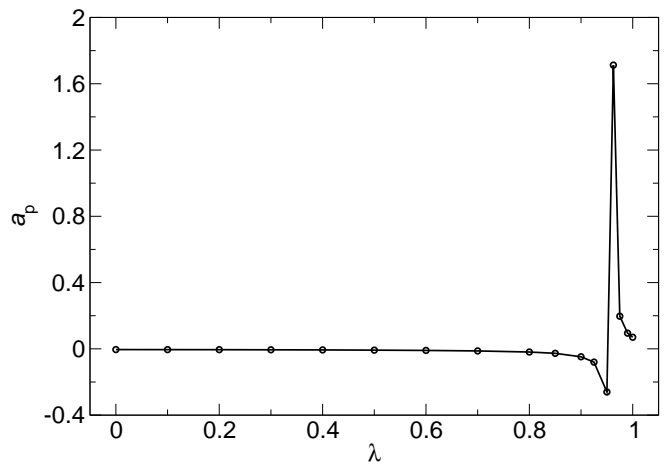}}
\caption{Variation of p-wave volume $a_{\rm p}$ ($10^8 \mbox{bohr}^3$) with reduced mass. See text.}
\label{fig:eps3}
\end{figure}

Furthermore, in the case of the singlet state characterised by the $X{}^1\Sigma^+$ molecular potential, the signs of the scattering volumes for the two isotopomers differ. To interpret this result we first compare the numbers of p-wave bound states of the two isotopomers. These can be obtained by applying Levinson's theorem in the same fashion as for s-waves. The accumulated p-wave phase shift, $\delta_{k,1}(R)$, satisfies the equation \cite{CAL67}:

\begin{equation} \label{eq7}
\frac{\displaystyle {\rm d}\delta_{k,1}(R)}{\displaystyle {\rm d}R} = - \frac{V(R)}{k} \left[\cos\delta_{k,1}(R) {\hat \jmath}(kR) - \sin\delta_{k,1}(R) {\hat n}(kR)\right]^2,
\end{equation}

\noindent where ${\hat \jmath}$ and ${\hat n}$ denote the Riccati-Bessel functions of order 1:
${\hat \jmath}(z)+i{\hat n}(z)=(iz^{-1}-1)\exp(-iz)$. 
The number of p-wave bound states supported by the $X{}^1\Sigma^+$ potential is changed from 82 for ${}^{23}\mbox{Na}{}^{85}\mbox{Rb}$ to 83 for ${}^{23}\mbox{Na}{}^{87}\mbox{Rb}$, demonstrating that with the increased reduced mass of the heavier isotope the interaction potential supports an extra bound level. In the case of the triplet state characterised by the $a{}^3\Sigma^+$ molecular potential, the numbers of p-wave bound states are the same, 23, for both isotopomers. The much larger value for ${}^{23}\mbox{Na}{}^{87}\mbox{Rb}$ suggests the possibility of a near zero energy bound state.

To find evidence for this we studied the dependence of the p-wave volume on the reduced mass of the NaRb molecule. We have defined a reduced mass, $\mu'$, which varies between that of the lighter isotopomer, $\mu_{\rm l}$, and that of the heavier one, $\mu_{\rm h}$:

\begin{equation}
\mu' = \mu_{\rm l} + \lambda (\mu_{\rm h}-\mu_{\rm l}),
\end{equation}

\noindent where $\lambda$ is in the range [0,1]. The  resonance in the variation of the scattering volume with reduced mass, shown in Fig.~\ref{fig:eps3}, suggests that the potential just supports a p-wave bound state for the heavier isotope. At resonance the parameter $\lambda\approx 0.9625$. The s-wave scattering length at the corresponding mass is 116 bohr and, as predicted by Dickinson \cite{DIC08}, is approximately twice the mean scattering length defined by Gribakin and Flambaum \cite{GRI93}, which is 55 bohr. Table \ref{tab:table1} shows that the values of the scattering volumes are close to those predicted by Dickinson's relation between the scattering length and the scattering volume \cite{DIC08} except near the resonance. In Ref.~\cite{DIC08}, the quantal and semiclassical values are in good agreement because the molecular potentials used for the calculations, do not support an additional p-wave bound state as does the improved potential of Ref.~\cite{PAS05}.

\section{Conclusion}

We calculated parameters that describe low energy collisions of Na and Rb atoms. We confirmed that the long range part of molecular potentials is sufficiently well characterised by the set of three van der Waals dispersion coefficients $C_6$, $C_8$ and $C_{10}$ to calculate the scattering data. We estimated temperatures below which only pure s-wave contributes to the scattering cross sections and showed that the inelastic spin change collisions dominate the s-wave elastic scattering in these systems. We found evidence for the existence of a near zero energy p-wave bound state. This is in accord with the semiclassical study made by Dickinson \cite{DIC08}.

\begin{acknowledgments}
We are pleased to thank Alex Dalgarno FRS for a useful discussion. HO acknowledges partial support of the Agence Nationale de la Recherche. Part of this work was done during his stay at the LASMEA, UMR CNRS-Universit\'e Blaise Pascal 6602, Aubi\`ere, France. MJJ thanks the Institute for Theoretical Atomic and Molecular Physics at the Harvard-Smithsonian Center for Astrophysics for supporting a research visit; the Institute is supported by a grant from the National Science Foundation. Partial support from the Chemical Sciences, Geosciences and Biosciences Division of the Office of Basic Energy Sciences, Office of Science, U.S. Department of Energy is acknowledged.
\end{acknowledgments}


\begin{thebibliography}{99}
\bibitem{WEI99} J. Weiner, V. S. Bagnato, S. Zilio, and P. S. Julienne, Rev. Mod. Phys. {\bf 70}, (1999) 1.
\bibitem{TRU01} A. G. Truscott, K. E. Stecker, W. I. McAlexander, G. B. Partridge, and R. G. Hulet, Science {\bf 291}, (2001) 2570.
\bibitem{SCH01} F. Schreck, L. Khaykovich, K. L. Corwin, G. Ferrari, T. Bourdel, J. Cubizolles, and C. Salomon, Phys. Rev. Lett. {\bf 87}, (2001) 080403.
\bibitem{ROA02} G. Roati, F. Riboli, G. Modugno, and M. Inguscio, Phys. Rev. Lett. {\bf 89}, (2002) 150403.
\bibitem{MOD02} G. Modugno, M. Modugno, F. Riboli, G. Roati, and M. Inguscio, Phys. Rev. Lett. {\bf 89}, (2002) 190404.
\bibitem{ESR97} B. D. Esry, C. H. Greene, J. P. Burke Jr., and J. L. Bohn, Phys. Rev. Lett. {\bf 78}, (1997) 3594.
\bibitem{BUR97} J. P. Burke Jr., J. L. Bohn, B. D. Esry, and C. H. Greene, Phys. Rev. A {\bf 55}, (1997), R2511. 
\bibitem{OUE03} H. Ouerdane, M. J. Jamieson, D. Vrinceanu and M. J. Cavagnero, J. Phys. B {\bf 36}, (2003) 4055.
\bibitem{OUE04} H. Ouerdane and M. J. Jamieson, Phys. Rev. A {\bf 70}, (2004) 022712.
\bibitem{WEI03} S. B. Weiss, M. Bhattacharya, and N. P. Bigelow, Phys. Rev. A {\bf 68}, (2003) 042708.
\bibitem{COT05} R. C\^{o}t\'{e}, R. Onofrio, and E. Timmermans, Phys. Rev. A {\bf 72}, (2005) 041605(R).
\bibitem{PAS05} A. Pashov, O. Docenko, M. Tamanis, R. Ferber, H. Knoeckel, and E. Tiemann, Phys. Rev. A {\bf 72}, (2005) 062505.
\bibitem{WEI04} S. B. Weiss, M. Bhattacharya, and N. P. Bigelow, Phys. Rev. A {\bf 69}, 049903(E) (2004). 
\bibitem{SMI65} B. M. Smirnov and M. I. Chibisov, Sov. Phys. JETP {\bf 21}, (1965) 624.
\bibitem{ZEM99} W. T. Zemke and W. C. Stwalley, J. Chem. Phys. {\bf 111}, (1999) 4962.
\bibitem{MAR96} M. Marinescu and A. Dalgarno, Zeitschrift Physik D - Atoms, Molecules and Clusters {\bf 36}, (1996) 239.
\bibitem{NISTJ} NIST Database at http://physics.nist.gov/PhysRefData.
\bibitem{STA85} J. M. Standard and P. R. Certain, J. Chem. Phys. {\bf 83}, (1985) 3002.
\bibitem{DER01} A. Derevianko, J. F. Babb, and A. Dalgarno, Phys. Rev. A {\bf 63}, (2001) 052704.
\bibitem{DER03} S. G. Porsev and A. Derevianko, J. Chem. Phys. {\bf 119}, (2003) 844.
\bibitem{MIT05} J. Mitroy and M. W. J. Bromley, Phys. Rev. A {\bf 71}, (2005) 042701.
\bibitem{DAL56} A. Dalgarno and R. McCarroll, Proc. Roy. Soc. London A {\bf 237}, (1956) 383.
\bibitem{CAL67} F. Calogero, \emph{Variable Phase Approach to Potential Scattering} (Academic Press, New York, 1967).
\bibitem{LEV49} N. Levinson, K. Dan. Vidensk. Selsk. Mat. Fys. Medd. {\bf 25}, (1949) 9.
\bibitem{LEV63} B. R. Levy and J. B. Keller, J. Math. Phys. {\bf 4}, (1963) 54.
\bibitem{HIN71} O. Hinckelmann and L. Spruch, Phys. Rev. A {\bf 3}, (1971) 642.
\bibitem{MOT65} N. F. Mott and H. S. W. Massey, \emph{The Theory of Atomic Collisions} (Oxford: Clarendon, 1965).
\bibitem{OUE04b} H. Ouerdane and M. J. Jamieson, J. Phys. B {\bf 37}, (2004) 3765.
\bibitem{GUT84} G. Guti\'errez, M. de Llano and W. C. Stwalley, Phys. Rev. B {\bf 29}, (1984) 5211.
\bibitem{DAL61} A. Dalgarno, Proceedings of the Royal Society of London A {\bf 262}, (1961) 132.
\bibitem{BLA48} J. M. Blatt, Phys. Rev. {\bf 74}, (1948) 92.
\bibitem{BLA49} J. M. Blatt and J. D. Jackson, Phys. Rev. {\bf 76}, (1949) 18.
\bibitem{BET49} H. A. Bethe, Phys. Rev. {\bf 76}, (1949) 38.
\bibitem{HAD02} Z. Hadzibabic, C. A. Stan, K. Dieckmann, S. Gupta, M. W. Zwierlein, A. G\"orlitz and W. Ketterle, Phys. Rev. Lett. {\bf 88}, (2002) 160401.
\bibitem{NIST} National Institute of Standards and Technology (NIST) www.nist.gov
\bibitem{DIC08} A. S. Dickinson, J. Phys. B {\bf 41}, (2008) 175302.
\bibitem{GRI93} G. F. Gribakin and V. V. Flambaum, Phys. Rev. A {\bf 48}, (1993) 546.
\end{thebibliography}
\end{document}